\documentclass[preprint,authoryear,12pt]{elsarticle}

\usepackage{amssymb}

\def\astrobj#1{#1}
\journal{New Astronomy}

\begin{document}

\begin{frontmatter}



\title{Close Binary System GO Cyg}


\author[ege]{B. Ula{\c s}}
\author[ege,iyt]{B. Kalomeni}
\author[ege]{V. Keskin}
\author[ege]{O. K\"ose}
\author[ege,cam]{K. Yakut \corref{kyv}}
\address[ege]{Department of Astronomy and Space Sciences, University of Ege, 35100, Bornova--{\.I}zmir, Turkey}
\address[iyt]{Department of Physics, {\.I}zmir Institute of Technology, Turkey}
\address[cam]{Institute of Astronomy, University of Cambridge, Madingley Road, Cambridge CB3 0HA, UK}
\cortext[kyv]{Visiting astronomer during the summer of 2011}

\begin{abstract}
In this study, we present long term  photometric variations of the close binary system \astrobj{GO Cyg}.
Modelling of the system shows that the primary is filling Roche lobe and the secondary of the system is almost filling its Roche lobe.
The physical parameters of the system are $M_1 = 3.0\pm0.2 M_{\odot}$, $M_2 = 1.3 \pm 0.1 M_{\odot}$,
$R_1 = 2.50\pm 0.12 R_{\odot}$, $R_2 = 1.75 \pm 0.09 R_{\odot}$, $L_1 = 64\pm 9 L_{\odot}$, $L_2 = 4.9 \pm 0.7 L_{\odot}$, and $a = 5.5 \pm 0.3 R_{\odot}$.
Our results show that \astrobj{GO Cyg} is the most massive system near contact binary (NCB).
Analysis of times of the minima  shows a sinusoidal variation with a period of  $92.3\pm0.5$ years due to a third body whose mass is less than 2.3$M_{\odot}$.
Finally a period variation rate of $-1.4\times10^{-9}$ d/yr has been determined using all available light curves.
\end{abstract}


\end{frontmatter}

\section{Introduction}

Studies of the evolution of late-type close binary systems reveal that the evolution of detached,
semi-detached and contact systems are closely related (Yakut \& Eggleton 2005, Eggleton 2010 and reference therein).
The more massive star in a detached binary system fills its Roche lobe first because it has shorter evolutionary timescale before its companion.
The system is semi-detached binary. In addition to nuclear evolution and mass loss, mass transfer
has a crucial role in driving a binary towards a contact phase of evolution. The observations of detached, contact and
semi-detached binaries are crucial to our further understanding of the evolution of close binary systems.

We therefore, include \astrobj{GO Cyg} (HD 196628, GSC 02694-00550, $V$=8$^m$.47, A0V) into our close binary stars observation programme
(see Ula{\c s} et al. 2011, K\"ose et al. 2011). The system is a $\beta$-Lyr type (short period 0$^{d}$.71)
binary system and observations of the binary cover eighty years.
Following its discovery by Schneller (1928) the system
has been extensively studied by many authors. (Payne-Gaposchkin, 1935; Pierce, 1939 and Popper, 1957).
Ovenden (1954); Mannino (1963); Rovithis et al. (1990); Sezer et al. (1993); Jassur (1997); Rovithis-Livanou et al. (1997);
 Edalati \& Atighi (1997);  Oh et al (2000), and Zabihinpoor et al. (2006) studied the system photometrically.
 Using different methods in analysis most studies agree with the primary filling its Roche lobe.
 Asymmetry in the secondary minimum have been discussed in previous studies (e.g. Edalati \& Atighi 1997, Zabihinpoor et al. 2006).
 Pearce (1933) found the mass function and mass ratio of $0.85$ for \astrobj{GO Cyg}. Later studies have reported
 higher mass ratios. Pribulla et al. (2009) examined the binary and classified it as a member of
 a group called {\it difficult binary stars}. In this group accurate radial velocities are not available. Pribulla et al. (2009)
 concluded that the temperature difference between the components makes the system a difficult candidate in determination
 of the ideal broadening function. A velocity value  of $v \approx 35$ km/s for a third body was given in Pribulla et al. (2009).

Period variation of \astrobj{GO Cyg} has been studied by many investigators. Period increase was discussed in a number of
studies (e.g. Sezer et al. 1985, Rovithis et al. 1997, Edalati \& Atighi 1997, Zabihinpoor et al. 2006). Jones et al. (1994)
reported a sinusoidal variation superimposed on a parabolic trend. Elkhateeb (2005) noted a period increase of $dP/dt=1.28\times 10^{-7}$
which is close to that of Oh et al.'s (2000) value of $1.51\times 10^{-7}$. Hall \& Louth (1990) discussed a magnetic cycle by studying
the period decrease between the year 1934 and 1984. Chochol et al. (2006) represented the $O-C$ curve by a sinusoidal fit
by using the Cracow database and their data. A third body with an orbital period of 90 years and a mass of 0.62$M_{\odot}$ has been proposed.

In the following sections, we present our new observations of \astrobj{GO Cyg}. We preformed photometric analysis, period variations 
and compare our results with its previously published works. All the available light curves were collected from the literature
and studied for various physical processes (e.g. magnetic activity, mass transfer, third light) and their variations.
The $O-C$ variation with recently obtained times of minima revealed the discrepancy between the results of the light curve
solution and period study of earlier studies. In this study, therefore, we investigate different possibilities that cause
period variation in order to reveal the most accurate structure and behavior of the components.
The physical parameters of the system are given with a discussion on the evolutionary status of the binary.

\section{New Observations}

The light variation and minima times of \astrobj{GO Cyg} obtained in the Bessel $B$, $V$, and $R$ bands in 16 nights between
June -- August 2007 and one night in April 2011. The observations carried out at T\"UB\.ITAK National Observatory (TUG) and
Ege University Observatory with the 40cm telescope using an Apogee CCD U47.
Comparison and check stars are selected as GSC 02694-00280 and GSC 02694-00733, respectively.
The total number of the points obtained during the observations are 3715 in $B$, 3726 in $V$, and 3698 in $R$ band.
IRAF (DIGIPHOT/APPHOT) packages are used in data reduction. Standard deviations of the data are estimated
as 0$^{\rm m}$.04, 0$^{\rm m}$.017, and 0$^{\rm m}$.015 for $B$, $V$, and $R$ bands, respectively.

In Fig.~1 we show the $B$, $V$, and $R$ light curves of \astrobj{GO Cyg}. In this study, we do not find the apparent
asymmetry in 0.6-0.7 orbital phase previously reported by Edalati \& Atighi (1997) and Zabihinpoor et al. (2006).
In data our reduction and analysis, we used the linear ephemeris described by Sezer et al. (1993).

\section{Eclipse Timings and Period Study}

Cester et al. (1979) reported an orbital increase of $Q=0.7\times 10^{-10}$ based on seven nights observational data obtained
between 1972-1975. Sezer et al. (1985) also reported an increase with a $Q$ value of $1.13\times 10^{-10}$ days.
Hall \& Louth (1990) consider the $O-C$ curve and split it in three region. The first and the third region of the curve showed sudden variation.
Therefore the authors analysed these regions under linear assumption while the second part was presented by a quadratic fit with a
period increase of $Q=1.28\times 10^{-10}$. The authors concluded that this behavior of the $O-C$ curve can be attributed to a
magnetic cycle. Jones et al. (1994) showed that the residuals of the parabolic fit show a sine-like variation with a period of 38.9 years.
A period increase was also discussed by Rovithis-Livaniou et al. (1997). A period with $Q=1.6\times 10^{-10}$ was noted by Edalati \& Atighi (1997).
Oh et al. (2000) also represented the $O-C$ curve by an upward parabola with $Q=1.47\times 10^{-10}$. Elkhateeb (2005) found that the period is
increasing with a value of $Q=1.26\times 10^{-10}$. The parabolic and third order polynomial fits were compared by Zabihinpoor et al. (2006).
The authors give the quadratic term as $0.935\times 10^{-10}$. Zabihinpoor et al. (2006) also discussed the inconsistency between geometric
configuration and period variation rate. Recently, Chochol et al. (2006) suggested the light time effect for the period variation.
The authors discussed that in a binary system where the primary fills its Roche lobe and loses mass a decrease in orbital period can be expected.

Recently we obtained two times of minima 24 54318.54864$\pm 0.00013$. and 24 55676.56220$\pm 0.00016$. The times of minima obtained in this study show that
the $O-C$ curve changes shape from early-assumed upward parabola to a sinusoidal variation that supports the previous discussion about
a third body in the system. The period variation is studied using a total of 194 data points obtained by photometric/CCD observations.
The times of minima are obtained from the literature (Kreiner et al., 2001 and Erkan et al., 2010) and those yielded by this study.
The weighted least-squares method is used in order to determine the orbital elements of the third body. Sinusoidal variation in the O-C curve,
where both the primary and the secondary minima follow the same trend suggests a light-time effect because of the
presence of a third component that can be represented by following formula (Irwin 1959, Kalomeni et al. 2007):

\begin{eqnarray}\nonumber
MinI&&= T_o +P_oE + \\
&& + \frac {a_{12} \sin i'}{c} \left[ \frac {1-{e'}^2}{1+{e'} \cos v'}\left( v' + \omega' \right) +{e'} \sin \omega' \right]
\end{eqnarray}
where $T_{\rm o}$ is the starting epoch for the primary minimum, $E$ is the integer
eclipse cycle number, $P_{\rm o}$ is the orbital period of the
eclipsing binary  $a_{12}$, $i'$ , $e'$, and $\omega'$
are the semi-major axis, inclination, eccentricity, and the
longitude of the periastron of eclipsing binary about the third body,
and $v'$ denotes the true anomaly of the position of the center of
mass. Time of periastron passage $T'$ and orbital period $P'$ are
the unknown parameters in Eq. (1).

Our result from our analysis are shown in Fig. 2. Fig. 2a shows the consistency between the observational and model prediction in the
assumption of a third body. Fig.~2b shows the residuals from a Sinusoidal variation. The orbital elements of a third component are
listed in Table 2. It can be clearly seen from the figure that any investigation of any detailed variation in $(\Delta T)_{II}$
points makes no sense. We have also investigated orbital period variation of the system with a different method from the $O-C$ analysis.
We have re--analysed the light curves from the available eighty years and we discussed further in Section 5 below. We conclude that if the
$O-C$ variation shows a downward parabola its period that is too long to determine from available times of minima.

\section{Light Curve Solution}

The light curve of the system has been analysed by numerous researchers. Ovenden (1954) obtained two-colour light curve and solved
them by using Russell's method. The author assumed that the primary dominates observed light. This results
in a reflection effect that makes difficult to identify the secondary in the spectrum. Asymmetry between maxima
discussed as an intrinsic variation. Mannino (1963) solved the photoelectric $B$ and $V$ light curves of the system with
the Russell--Merill method. Rovithis et al. (1990) analysed the light curves and estimated the geometric elements by using
frequency domain techniques. The authors reported no difference between the level of maxima. The $BV$ light curve combined
with Holmgren's radial velocity curve was solved by Sezer et al. (1993) by using Wilson--Devinney (WD) method.
The result indicated a semi--detached configuration where the primary is filling its Roche lobe.
A standard iterative optimization technique was used by Jassur (1997) to solve the $UBV$ light curves. Rovithis-Livaniou et al. (1997)
determined the absolute parameters and the geometrical elements by applying the Wood's model. Edalati \& Atighi (1997)
compared some parameters of their solution with previous works and confirmed that the system's geometrical configuration
is a reverse Algol. Oh et al. (2000) discussed that the system is at poor thermal contact phase of the thermal relaxation oscillation.
Recently, Zabihinpoor et al. (2006) analyzed the light curve and suggested new observations to uncover the discrepancy between the
suggested geometric shape and the orbital period variation.

The shapes of the radial velocity curves of the system are controversial and no reliable spectroscopic mass ratio exist in the literature.
In this study, therefore, we started to the solution by searching the appropriate photometric mass ratio. $q$ values between $0.25$ and $0.65$
are investigated on the $V$ light curve by increasing the value by $0.05$. We reached the minimum residual when $q=0.45$ which is then taken as
an initial value for our simultaneous solution. The uncertainties of the spectral types also required to search for a suitable temperature for the primary
component using the light curve. $T_{\rm h}$=10350 K turned out to be a suitable mean temperature of the hot component and has been used in other studies.
Simultaneous solutions obtained with {\sc Phoebe} (Pr\~{s}a \& Zwitter 2005), which uses the WD code (Wilson \& Devinney 1971), was applied to
our observations (476 points in $B$ and $V$, and 474 in $R$). The gravity darkening coefficients $g_1$ and $g_2$ are obtained from
von Zeipel (1924) and Lucy (1967). The albedos $A_1$ and $A_2$ are adopted from Rucinski (1969). The logarithmic limb-darkening law is used with
coefficients adopted from van Hamme (1993) for a solar composition (Table~2).
The adjustable parameters are orbital inclination \textit{i}, temperature of secondary component $T_2$, surface potential of
secondary component $\Omega_2$, luminosity $L_1$, and mass ratio $q$. The analysis results are summarized in Table~2.
The computed light curves are shown with solid lines in Fig.~1.

All available light curves of the system between 1936--2007 are also collected from the literature and analysed separately (Table~3).
These light curves are solved by using the initial values that are determined in this study. In addition, light contribution of
the third body is set as a fixed parameter since no variation is expected in a period of eighty years.
The results are shown in Fig. 3 and listed in Table~4. All available data are provided in Table~5.

Some light curves (LC8, LC12, LC15) show slight asymmetry in the secondary minimum while it is not detected in the others
(LC3, LC7, LC14). In this study we also investigate any evidence of a magnetic activity as it was discussed in earlier studies (Hall \& Louth 1990).
Either physical structure of the stars or the shape of the light curves did not let us to analyse the curves with spotted model assumption.
Some light curves (LC8, LC15), however, can be represented theoretically by a hot surface on the cooler companion.
The presence of a hot region, that may be attributed to a mass transfer other than a magnetic activity, could not be proved with the long-term data.

\section{Discussion and Conclusion}

Long term photometric light and period variation of the close binary system \astrobj{GO Cyg} are studied in detail.
The physical parameters we have determined and listed in Table~6.
Because of poor quality data in the previous studies the $O-C$ curve was inferred to be parabolic one.
In this study with our data we show that the system has a third body with a 92.3 years orbital period. Since the primary component
filled its Roche lobe we searched for a clue for the mass transfer by a technique other than the $O-C$ analysis.
Separate solutions of six available light curves starting from 1950 to present indicate a period decrease.
The period change variation vs. years shows a downward parabola. This can be considered
as a mass transfer from the more massive companion to the less massive one.
This solution yields the amount of the period decrease as $-1.4\times 10^{-9}$  d/yr with a mass transfer rate
of $1.5\times 10^{-9}$ {M$_{\odot}$/yr}. In addition, the period of this parabolic variation
is too long to detect in the $O-C$ curve constructed with the available data.

The system has been studied spectroscopically but no accurate radial velocity
study of it is available in the literature. In systems like GO Cyg the luminous primary star makes
it difficult to treat the system as a double-lined binary. Determination of the physical parameters requires the
knowledge of accurate mass ratio and the mass of the primary component. The best result obtained by a \textit{q}-search
technique in the light curve analysis.
In what follows solution to allowed the mass ratio to vary allowing with other parameters when the solve for an orbital the light curves.
Our solutions all have similar results for all nine light curves. Our results are given in Table~2-3.
Spectral studies of NCB systems (e.g. GO Cyg) are quite difficult because of their nature.
The mass of the primary, therefore, estimated according to their colors, spectral types \textit{etc}.
In this study, we used recently published astrophysical data of well known stars (e.g., Torres et al., 2010, Yakut \& Eggleton, 2005, Drilling \& Landolt 2000)
to estimate the primary's (the massive and lumonious one) mass.
By studying stars with similar luminosities and spectral types, the mass of the primary is assumed to be 3.0 M$_{\odot}$.
This is also consistent with the values given in the literature.
The physical parameters of the system is given in Table~6, the results are consistent with the similar
systems on the $M-R$, $M-L$ and the Hertzsprung--Russell diagrams given by Yakut \& Eggleton (2005).
Our results shows that GO Cyg has the most massive components among the known NCB systems.
We collected physical parameters of the NCB systems whose primary components are relatively massive (Table~7).
Mass-luminosity diagram of binaries listed in Table~7 is shown in Fig~4.
The locations of GO Cyg A and B in the $M-L$ diagram are consistent with the other NCB systems.

Observational results of semi-detached systems show that while in some cases the primary
component fills its Roche lobe (GO Cyg) in other cases the secondary fills
its Roche lobe (V836 Cyg, Yakut et al., 2005). These differences are a sign for the evolutionary stage of the
binary system (for details we refer Yakut \& Eggleton 2005, Eggleton 2006, Eggleton 2010).
The orbital, geometrical, and physical parameters of GO Cyg presented in this study
indicate the Roche lobe filling star is the primary (the massive and the hotter one).
We show that the primary may even transfer mass with at a low rate.
Contrary to the very low mass stars, mass loss rate due to the
magnetic stellar winds can be expected since the convective layer is small in the systems with intermediate/low
mass components (e.g. GO Cyg). The results indicate the system GO Cyg evolves under the proximity effect,
low rate mass transfer between the components ($\dot{M}=1.5\times10^{-9}$ M$_\odot$/yr) and the third body can also remove angular momentum from the binary orbit.

\section*{Acknowledgments}
This study was supported by the Turkish Scientific
and Research Council (T\"UB\.ITAK 109T047 and 111T270) and Ege University Research Fund.
KY+VK acknowledges support by the Turkish Academy of Sciences (T{\"U}BA).
We thank to an anonymous referee, J.J. Eldridge and E.R. Pek\"unl\"u for their valuable comments and suggestions.

\begin{figure}
\includegraphics[scale=0.50]{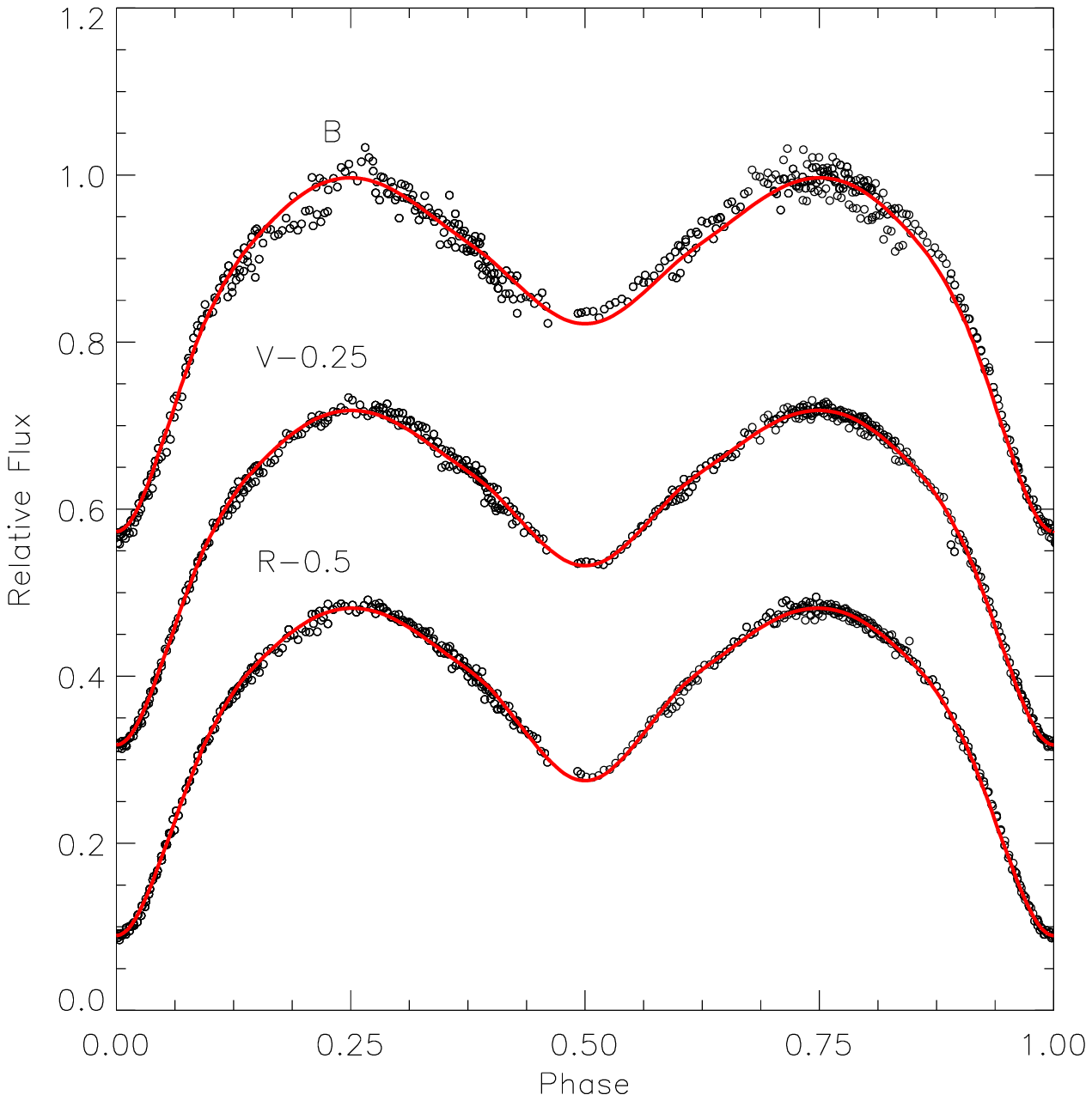}\\
\caption{The observed and the computed (solid line) light curves of the system GO Cyg.
The light curves in $V$ and $R$ bands are moved by a value of $0.25$ and $0.5$, respectively, for the sake of comparison.}\label{fig1}
\end{figure}

\begin{figure}
\includegraphics{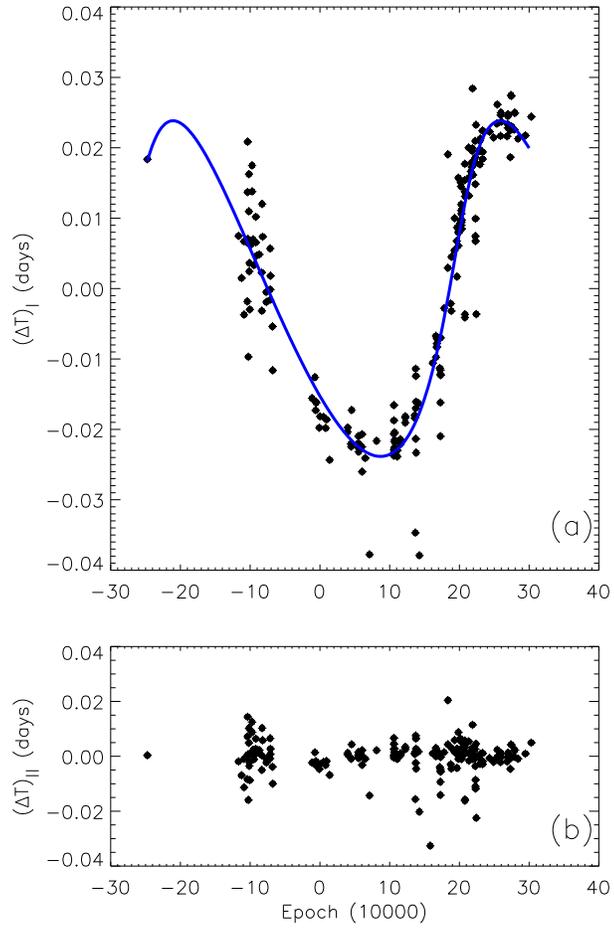}
\caption{(a) Residuals for the times of minimum light of GO Cyg. The solid line is obtained with the assumption of sine-like variation. (b) The difference between the observations and the computed sinusoidal curve.}\label{fig2}
\end{figure}

\begin{figure*}
\includegraphics[scale=0.78]{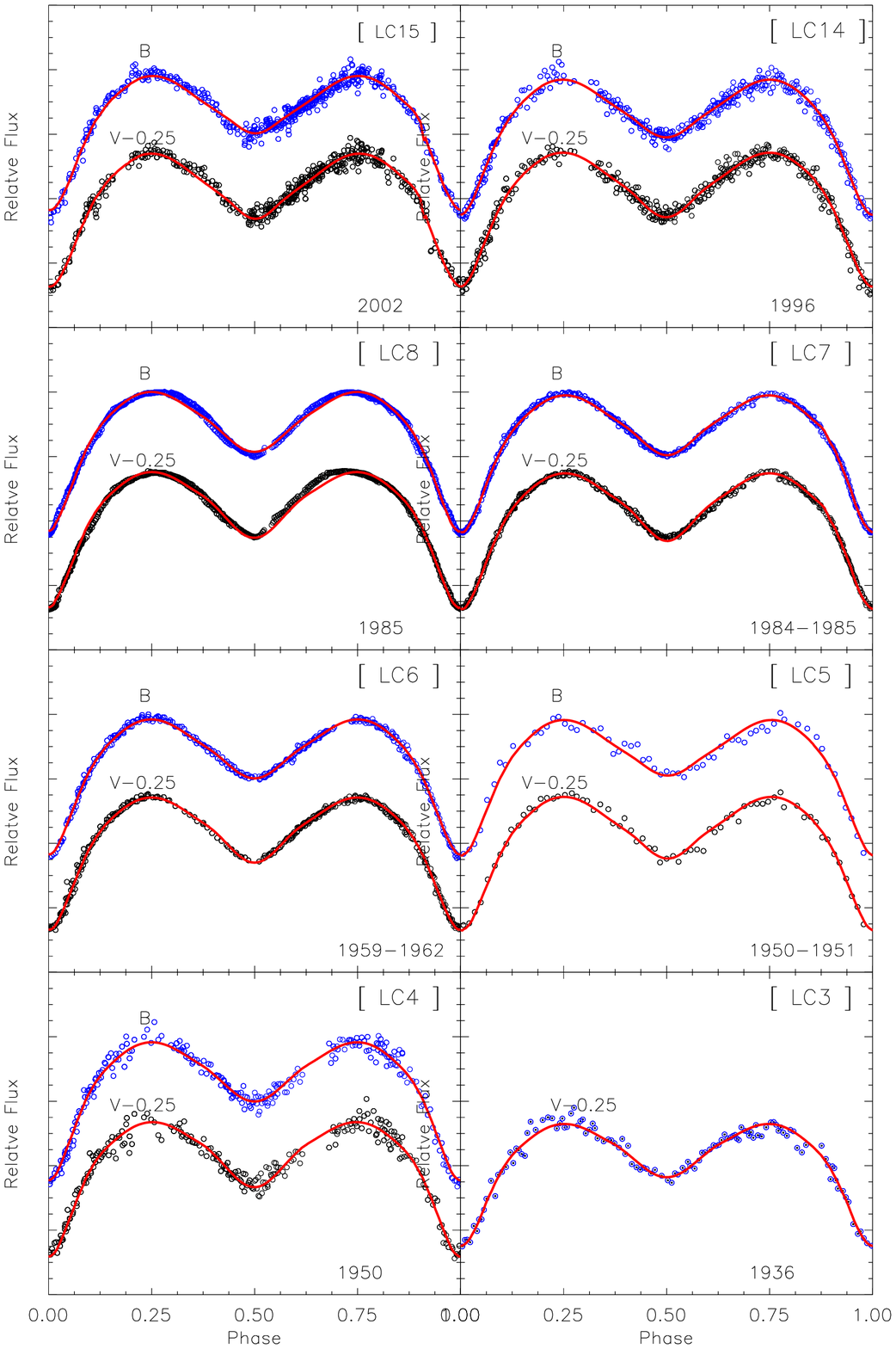}\\
\caption{All available light curves (phase-intensity) of the system between 1936 and 2002. The theoretical curves (solid lines) are drawn using the results given in Table~6. At the right bottom of each panel observation years are given.}\label{fig3}
\end{figure*}


\begin{figure}
\includegraphics{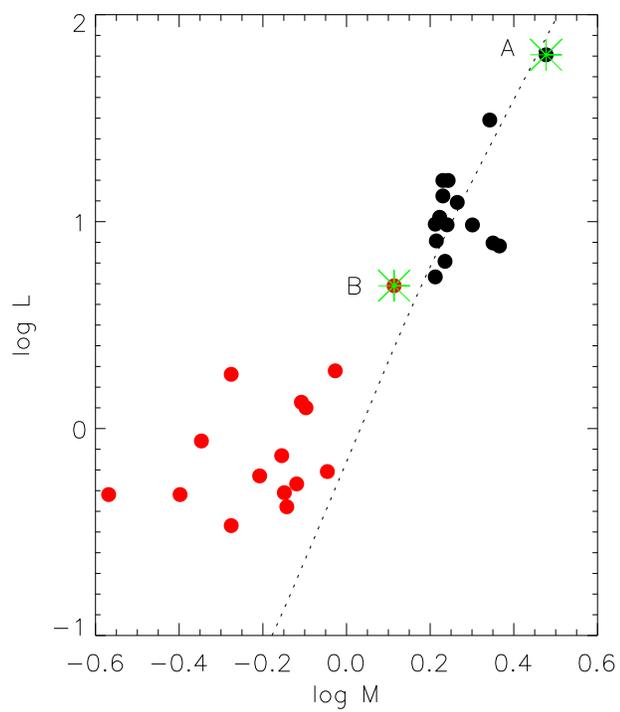}
\caption{Plot of the $M-L$ plane of the some NCB systems. The ZAMS line is taken from Pols et al. (1995).} \label{fig4}
\end{figure}

\begin{table}
\caption{Results of the period analysis and the orbital elements of the third body. The
standard errors, 1$\sigma$, are given in parentheses.} \label{tab1}
\begin{tabular}{lll}
\hline
Parameter            &Unit               & Value                       \\
\hline $T_o$         & [HJD]             & 2433930.4283(7)             \\
$P_o$                & [day]             & 0.717764585(15)                 \\
$P'$                 & [year]            & 92.3(5)                        \\
$T'$                 & [HJD]             & 2414756(300)                      \\
$e'$                 &                   & 0.46(1)                          \\
$\omega'$            & [$^\circ$]        & 20.3 (2.2)                              \\
$a_{12} \sin i'$     & [AU]              & 4.57 (4)                               \\
$f(m)$               & [M$_{\odot}$]     & 0.0112(5)                                  \\
$m_{3;i'=20^\circ}$  & [M$_{\odot}$]     & 2.30                                        \\
$m_{3;i'=90^\circ}$  & [M$_{\odot}$]     & 0.65                                     \\
\hline
\end{tabular}
\end{table}

\begin{table}
\caption{The photometric elements and their formal
1$\sigma$ errors of GO Cyg. See text for details.}
\label{tab2}
\begin{tabular}{lll}
\hline
Parameter                                   & Value      \\
\hline
Geometric parameters:                       &            \\
$i$ ${({^\circ})}$                          & 75.67(3)   \\
$\Omega _{1}$                               & 2.735      \\
$\Omega _{2}$                               & 2.686(4)   \\
$q$                                     & 0.428(9)   \\
Fractional radius of primary                & 0.4542(2) \\
Fractional radius of secondary              & 0.3189(12) \\
Radiative parameters:                       &            \\
$T_1$ (K)                               & 10350   \\
$T_2$ (K)                                   & 6490(90)  \\
Albedo $A_1$                            & 1.0        \\
Albedo $A_2$                            & 0.760      \\
Gravity brightening $g_1$               & 1.0        \\
Gravity brightening $g_2$               & 0.32       \\
Limb darkening  $x_1, x_2$                   &            \\
$x_1B$                                      & 0.693      \\
$x_2B$                                      & 0.812      \\
$x_1V$                                      & 0.593      \\
$x_2V$                                      & 0.721      \\
$x_1R$                                      & 0.484      \\
$x_2R$                                      & 0.628      \\
Luminosity ratio:$\frac{L_1}{L_1 +L_2+l_3}$(\%) &   \\
$B$                                         & 94(5)          \\
$V$                                         & 90(4)        \\
$R$                                         & 87(4)         \\
Luminosity ratio:$\frac{l_3}{L_1 +L_2+l_3}$(\%) &   \\
$B$                                         & 0.39         \\
$V$                                         & 0.37         \\
$R$                                         & 0.38         \\
\hline
\end{tabular}
\end{table}


\begin{table*}
\scriptsize
\caption{Available light curves of GO Cyg that are collected from the literature. JD$^*$ refers to the time interval of data taken. In data availability column (Data), "Yes" and "No" is shortened by the letter Y and N, respectively. Photoelectric observations are abbreviated by "pe" while "pg" and "ccd" refers to the photographic and CCD observations. Light curves are LC1: Payne-Gaposchkin (1935), LC2: Liau (1935), LC3: Pierce (1939), LC4: Popper (1957), LC5: Ovenden (1954), LC6: Mannino (1963), LC7: Sezer et al. (1993), LC8: Rovithis et al. (1990), LC9: Oprescu et al. (1996), LC10: Jassur (1997), LC11: Rovithis-Livaniou et al. (1997), LC12: Edalati and Atighi (1997), LC13: Vukasovi\'{c} 1997, LC14:	Oh et al. (2000), LC15: Zabihinpoor et al. (2006), LC16: This study.}
\label{tab3}
\begin{tabular}{llllcllc}
\hline
ID	&	Year	&	JD$^*$(2400000+)	&	Filters	&	Type	&	Comparison(s)	&	N$_{\textrm{points}}$	& Data \\
\hline
LC1	&	1934	&		&blue, red		&pg&	BD +36 4150	&		&	N	\\
	&		&		&		&		&	BD +34 4098	&		&		\\
	&		&		&		&		&	BD +35 4197	&		&		\\
LC2	&	1935	&		&		&		&			&		&	N	\\
LC3	&	1936	&		&	V	&	vis	&	BD +35 4188	&	122	&	Y 	\\
LC4	&	1950	&33478.8-33498.0		&	B,V	&	pe	&	HD 196771	&B:261, V:261	&	Y	\\
LC5	&	1950-51	&		&	B,V	&	pe	&	BD +35 4197	&B:64, V:68		&	Y 	\\
	&		&		&		&		&	BD +34 4098	&		&		\\
LC6	&	1959-62	&36782.4-37910.5		&	B, V	&	pe	&	BD +35 4197	&B:333, V:353		&	Y	\\
	&		&		&		&		&	BD +34 4098	&		&		\\
LC7	&	1984-85	&45866.4-46348.3		&	B,V	&	pe	&	HD 197 292	&B:416, V:414	&	Y	\\
	&		&		&		&		&	HD 197 346	&		&		\\
LC8	&	1985	&46264.3-46329.4		&	B,V	&	pe	&	BD +35 4180	&B:631, V:633	&	Y	\\
	&		&		&		&		&	BD +34 4098	&		&		\\
LC9	&	1989-92	&		&	B,V	&	pe	&	BD +35 4197	&		&	N	\\
	&		&		&		&		&	BD +34 4098	&		&		\\
LC10	&	1992	&		&	U,B,V	&	pe	&	BD +35 4180	&		&	N	\\
	&		&		&		&		&	BD +34 4098	&		&		\\
LC11	&	1993-94	&		&	B,V	&	pe	&	BD +35 4197	&		&	N 	\\
	&		&		&		&		&	BD +34 4098	&		&		\\
LC12	&	1995	&		&	U,B,V	&	pe	&	BD +35 4180	&		&	N	\\
	&		&		&		&		&	BD +34 4098	&		&		\\
LC13	&	1996	&		&	U,B,V	&	pe	&	SAO 70314	&		&	N	\\
LC14	&	1996	&50366.0-50436.0		&	B,V	&	pe	&	BD +35 4197	&B:398, V:397	&	Y	\\
	&		&		&		&		&	BD +34 4098	&		&		\\
LC15	&	2002	&		&	B,V	&	pe	&	HD 197292	&B:545, V:521		&	Y	\\
	&		&		&		&		&	HD 197346	&		&		\\
LC16	&	2007	&54361.5-54321.6		&	B, V, R	&	ccd	&	GSC 02694-00280	&B:3711, V:3722,	&	Y	\\
	&		&		&		&		&	GSC 02694-00733	&	 R:3694		&		\\
\hline
\end{tabular}
\end{table*}

\begin{table*}
\small
\caption{The photometric parameters and 1$\sigma$ errors obtained from the solution of all available light curves. See text for details.}
\label{tab4}
\begin{tabular}{llllll}
\hline
Parameter	&	LC3	&LC4	&LC5	&	LC6	&LC7		    		       \\   
\hline												     
$i$ ${({^\circ})}$ 		&	75.2(1.5)&77.1(3)   &73.42(1.03)&	75.57(8)	     &76.61(3)\\   
$q$		   		&	0.371(26)&0.435(5)  &0.443(17)  &0.425(2) 		     &0.457(1)\\   
$T_1$ (K) 	   		&	10350	 &10350	    &10350      &	10350		     &10350   \\   
$T_2$ (K) 	   		&	6111(240)&6667(64)  &6516(144)  &	6721(30)	     &6743(18)\\   
$\Omega _{1}$	   		&	2.616	 &2.764	    &2.749      &	2.729		     &2.786   \\   
$\Omega _{2}$	   		&	2.603(65)&2.722(18) &2.627(18)  &	2.662(3)	     &2.702(2) \\   
$(\frac{L_1}{L_1 +L_2})_{B}$	&-		 &0.934(47) &0.932(101) &0.927(18)		     & 0.913(13)\\   
~~~~~~~~~~$_{V}$		&	0.938(124)   &0.896(53)	&0.895(122)&0.892(21)	    	     & 0.877(14)\\   
~~~~~~~~~~$_{R}$		&-	&-	&-   &	-				    	     & -	 \\   
$r_1$				&	0.4676(65)	&0.4512(11)	&0.4528(34)	&0.4549(3)	&0.4490(2)	\\   
$r_2$				&	0.2965(268)	&0.3184(46)	&0.3420(151)	&0.3238(16)	&0.3352(13)	\\   
\cline{2-5}	 										    
				 &	 LC8	 &	 LC14	 &	 LC15	&{\bf{LC16}} &	\\
\cline{2-5}	 
$i$ ${({^\circ})}$ 		 &77.02(2)	 &74.2(1)	 &74.5(1)     &75.67(3)     &	\\
$q$		   		 &0.424(2)	 &0.457(2)	 &0.461(2)    & 0.428(9)    &	\\
$T_1$ (K) 	   		 &10350 	 &10350 	 &10350       & 10350 	    &	\\
$T_2$ (K) 	   		 &6688(23)	 &6651(50)	 &6709(35)    & 6478(262)   &	\\
$\Omega _{1}$	   		 &2.713 	 &2.785 	 &2.804       & 2.735 	    &	\\
$\Omega _{2}$	   		 &2.711(6)	 &2.674(3)	 &2.612(3)    & 2.686(4)    &	\\
$(\frac{L_1}{L_1 +L_2})_{B}$	 &0.913(13)	 &0.933(15)	 &0.922(32)   & 0.941()59   &	\\
~~~~~~~~~~$_{V}$			 &0.906(16)	 & 0.885(36)	 &0.862(30)   & 0.908(48)   &	 \\
~~~~~~~~~~$_{R}$			 &-		 & -		 & -	      & 0.875(46)   &	 \\
$r_1$				 & 0.4567(4)	 & 0.4490(3)	 & 0.4472(3)  &  0.4542(2)  &	\\
$r_2$				& 0.3110(21)	 & 0.3405(15)	 & 0.3680(22) &  0.3189(12) &	\\

\end{tabular}
\end{table*}

\begin{table}
\caption{Available light curves data. Phases are given for the light curves LC3, LC5, and LC15 since JDs are not provided. All data for 9 data sets can be found electronically at CDS.} \label{tab5}
\begin{tabular}{cccc}
\hline
Data Set& Filter         &JD/Phase           & Magnitude  \\
\hline
LC6      &  B            &  2436782.3973   & 0.156        \\
LC6      &  B            &  2436782.4034   & 0.158        \\
LC6      &  B            &  2436782.4117   & 0.173        \\
LC6      &  B            &  2436782.4184   & 0.184        \\
LC6      &  B            &  2436782.4198   & 0.187        \\
\vdots   & \vdots        &  \vdots         & \vdots       \\
\hline
\end{tabular}
\end{table}

\begin{table}
\begin{center}
\caption{Absolute parameters of GO Cyg. The standard errors
1$\sigma$ in the last digit are given in
parentheses.}\label{tab6}
\begin{tabular}{llll}
\hline
Parameter                                        &Unit                      & Pr.           & Sec.   \\
\hline
Mass (M)                                         &$\rm{M_{\odot}}$      & $3.0(2)$            & $1.3(1)$      \\
Radius (R)                                       &$\rm{R_{\odot}}$      & $2.50(12)$           & $1.75(9)$      \\
Temperature (T$_{\rm eff}$)                      &$\rm{K}$              & $10350$              & $6490$    \\
Luminosity (L)                                   &$\rm{L_{\odot}} $     & $64(9)$           & $4.9(7)$      \\
Absolute bolometric magnitude (M$_b$)            &mag                   & 0.23             & 3.03            \\
Period change rate ($\dot{P}$)             &d/yr                 &~~~~~~~~~~~~~$-1.4\times10^{-9}$ &      \\
Mass transfer ratio ($\dot{M}$)            &M$_\odot$/yr         &~~~~~~~~~~~~~~~$1.5\times10^{-9}$ &      \\
Seperation between stars ($a$)                &$\rm{R_{\odot}}$  &~~~~~~~~~~~~~~~~~5.5(3) &      \\
\hline
\end{tabular}
\end{center}
\end{table}

\begin{table}
\caption{Physical parameters of some well known massive ($M > 1.6$ M$_{\odot}$) NCB systems. The data is taken from Yakut \& Eggleton (2005) except
RZ Dra	(Erdem et al., 2011), RU UMi (Lee et al., 2008), GW Gem	(Lee et al., 2009), EE Aqr	(Wronka et al. 2010), KQ Gem	(Zhang 2010).
} \label{tab7}
\begin{tabular}{lllllll}
Name	&	Sp.T&	P(d)	& M$_1$ 	& M$_2$ 	& $\log L_1$ & $\log L_2$\\	
\hline
\astrobj{RZ Dra}	&	A5	&	0.5509	&	1.63	&	0.70	&	0.988	&	-0.131	\\
\astrobj{RT Scl}	&	F0	&	0.5116	&	1.63	&	0.71	&	0.733	&	-0.310	\\
\astrobj{RV Crv}	&	F3	&	0.7473	&	1.64	&	0.45	&	0.907	&	-0.060	\\
\astrobj{AG Vir}	&	A8	&	0.6427	&	1.67	&	0.53	&	1.021	&	0.262	\\
\astrobj{DO Cas}	&	A7	&	0.6847	&	1.70	&	0.53	&	1.124	&	-0.469	\\
\astrobj{KQ Gem}	&	F5	&	0.4080	&	1.70	&	0.40	&	1.199	&	-0.319	\\
\astrobj{SW Lyn}	&	F2	&	0.6441	&	1.72	&	0.90	&	0.808	&	-0.208	\\
\astrobj{GW Gem}	&	A7	&	0.6594	&	1.74	&	0.80	&	0.985	&	0.100	\\
\astrobj{FO Vir}	&	A8	&	0.7756	&	1.75	&	0.27	&	1.199	&	-0.319	\\
\astrobj{YY Cet}	&	A8	&	0.7905	&	1.84	&	0.94	&	1.093	&	0.279	\\
\astrobj{RS Ind}	&	A9	&	0.6241	&	2.00	&	0.62	&	0.984	&	-0.229	\\
\astrobj{V836 Cyg}&	B9.5&	0.6534	&	2.20	&	0.78	&	1.491	&	0.127	\\
\astrobj{EE Aqr}	&	A9.5&	0.5090	&	2.24	&	0.72	&	0.897	&	-0.378	\\
\astrobj{RU UMi}	&	A9	&	0.5249	&	2.32	&	0.76	&	0.883	&	-0.268	\\
\astrobj{GO Cyg}	&	B8	&	0.7178	&	3.00	&	1.30	&	1.806	&	0.690	\\
\hline
\end{tabular}
\end{table}


\begin{thebibliography}{}
\bibitem[\protect\citeauthoryear{Cester et al.}{1979}]{1979Acta Astron.....29..433C} Cester B., Giuricin G., Mardirossian F., Mezzetti M., 1979, Acta Astron., 29, 433
\bibitem[\protect\citeauthoryear{Chochol et al.}{2006}]{2006ApSS.304...93C} Chochol D., et al., 2006, Ap\&SS, 304, 93
\bibitem[Drilling \& Landolt(2000)]{2000asqu.book..381D} Drilling, J.~S., \& Landolt, A.~U.\ 2000, Allen's Astrophysical Quantities, 381
\bibitem[\protect\citeauthoryear{Edalati and Atighi}{1997}]{1997ApSS...253..107}  Edalati M. T. \& Atighi M., 1997, Ap\&SS, 253, 107
\bibitem[\protect\citeauthoryear{Eggleton}{2006}]{2006epbm.book.....E} Eggleton, P. 2006, Evolutionary Processes in Binary and Multiple Stars, Cambridge Astrophysics Series No. 40
\bibitem[\protect\citeauthoryear{Eggleton}{2010}]{2010NewAR..54...45E} Eggleton P.~P., 2010, NewAR, 54, 45
\bibitem[\protect\citeauthoryear{Elkhateeb}{2005}]{2005JKAS...38...13E} Elkhateeb M.~M., 2005, JKAS, 38, 13
\bibitem[\protect\citeauthoryear{Erdem et al.}{2011}]{2011NewA..16...6E} Erdem, A., Zola, S., Winiarski, M. 2011, NewA, 16, 6
\bibitem[\protect\citeauthoryear{Erkan et al.}{2010}]{2010IBVS.5924....1E} Erkan N., Erdem A., Ak{\i}n T., Ali\c{c}avu\c{s} F., Soydugan F., 2010, IBVS, 5924, 1
\bibitem[\protect\citeauthoryear{Hall \& Louth}{1990}]{1990JApA...11..271H} Hall D.~S., Louth H., 1990, JApA, 11, 271
\bibitem[\protect\citeauthoryear{Irwin}{1959}]{1959AJ.....64..149I} Irwin J.~B., 1959, AJ, 64, 149
\bibitem[\protect\citeauthoryear{Jassur}{1997}]{1997ApSS.249..111J} Jassur D.~M.~Z., 1997, Ap\&SS, 249, 111
\bibitem[\protect\citeauthoryear{Jones et al.}{1994}]{1994IAPPP..54...34J} Jones R.~A., Snyder L., Frey G., Dalmau F.~J., Aloy J., Bonvehi L., 1994, IAPPP, 54, 34
\bibitem[\protect\citeauthoryear{Kalomeni et al.}{2007}]{2007AJ....134..642K} Kalomeni B., Yakut K., Keskin V., De{\u g}irmenci {\"O}.~L., Ula{\c s} B., K{\"o}se O., 2007, AJ, 134, 642
\bibitem[\protect\citeauthoryear{K{\"o}se et al.}{2011}]{2011AN....332..626K} K{\"o}se O., Kalomeni B., Keskin V., Ula{\c s} B., Yakut K., 2011, AN, 332, 626
\bibitem[\protect\citeauthoryear{Kreiner, Kim, \& Nha}{2001}]{2001aocd.book.....K} Kreiner J.~M., Kim C.-H., Nha I.-S., 2001, aocd.book
\bibitem[\protect\citeauthoryear{Lee et al.}{2008}]{2008PASP..120..720L} Lee J.~W., Kim C.-H., Kim S.-L., Lee C.-U., Han W., Koch R.~H., 2008, PASP, 120, 720
\bibitem[\protect\citeauthoryear{Lee et al.}{2009}]{2009PASP..121..104L} Lee J.~W., Kim S.-L., Lee C.-U., Kim H.-I., Park J.-H., Park S.-R., Koch R.~H., 2009, PASP, 121, 104
\bibitem[\protect\citeauthoryear{Liau}{1935}]{1935POLyo...1.....L} Liau S.~P., 1935, POLyo, 1
\bibitem[\protect\citeauthoryear{Lucy}{1967}]{lu} Lucy, L. B., 1967, Z. Astrophys., 65, 89
\bibitem[\protect\citeauthoryear{Mannino}{1963}]{1963MmSAI..34..191M} Mannino G., 1963, MmSAI, 34, 191
\bibitem[\protect\citeauthoryear{Oh et al.}{2000}]{2000ApSS.271..303O} Oh K.-D., Kang Y.~W., Ra K.~S., Park H.~S., 2000, Ap\&SS, 271, 303
\bibitem[\protect\citeauthoryear{Oprescu et al.}{1996}]{1996RoAJ....6..119O} Oprescu G., Dumitrescu A., Suran M.~D., Rovithis P., Rovithis-Livaniou H., 1996, RoAJ, 6, 119
\bibitem[\protect\citeauthoryear{Ovenden}{1954}]{1954MNRAS.114..569O} Ovenden M.~W., 1954, MNRAS, 114, 569
\bibitem[\protect\citeauthoryear{Payne-Gaposchkin}{1935}]{1935BHarO.898....3P} Payne-Gaposchkin C., 1935, BHarO, 898, 3
\bibitem[\protect\citeauthoryear{Pearce}{1933}]{1933JRASC..27...62P} Pearce J.~A., 1933, JRASC, 27, 62
\bibitem[\protect\citeauthoryear{Pierce}{1939}]{1939AJ.....48..113P} Pierce N.~L., 1939, AJ, 48, 113
\bibitem[\protect\citeauthoryear{Pols et al.}{1995}]{1995MNRAS.274..964P} Pols O.~R., Tout C.~A., Eggleton P.~P., Han Z., 1995, MNRAS, 274, 964
\bibitem[\protect\citeauthoryear{Popper}{1957}]{1957ApJS....3..107P} Popper D.~M., 1957, APJS, 3, 107
\bibitem[\protect\citeauthoryear{Pribulla et al.}{2009}]{2009AJ....137.3655P} Pribulla T., et al., 2009, AJ, 137, 3655
\bibitem[\protect\citeauthoryear{Prsa and Zwitter}{2005}]{pr} Pr\v{s}a, A., Zwitter, T., 2005, ApJ, 628, 426
\bibitem[\protect\citeauthoryear{Rovithis, Rovithis-Livaniou, \& Niarchos}{1990}]{1990A&AS...83...41R} Rovithis P., Rovithis-Livaniou H., Niarchos P.~G., 1990, A\&AS, 83, 41
\bibitem[\protect\citeauthoryear{Rovithis-Livaniou et al.}{1997}]{1997A&A...327.1017R} Rovithis-Livaniou H., Rovithis P., Oprescu G., Dumitrescu A., Suran M.~D., 1997, A\&A, 327, 1017
\bibitem[\protect\citeauthoryear{Rucinski}{1969}]{ru}Rucinski, S. M., 1969, Acta Astron., 19, 245
\bibitem[\protect\citeauthoryear{Schneller}{1928}]{sch} Schneller, H., 1928, AN, 235, 85
\bibitem[\protect\citeauthoryear{Sezer, Gulmen, \& Gudur}{1985}]{1985IBVS.2743....1S} Sezer C., G\"{u}lmen O., G\"{u}d\"{u}r N., 1985, IBVS, 2743, 1
\bibitem[\protect\citeauthoryear{Sezer, Gulmen, \& Gudur}{1993}]{1993ApSS.203..121S} Sezer C., G\"{u}lmen O., G\"{u}d\"{u}r N., 1993, Ap\&SS, 203, 121
\bibitem[\protect\citeauthoryear{Torres, Andersen, \& Gim{\'e}nez}{2010}]{2010A&ARv..18...67T} Torres G., Andersen J., Gim{\'e}nez A., 2010, A\&ARv, 18, 67
\bibitem[\protect\citeauthoryear{Ulas et al.}{2011}]{2011arXiv1107.0277U} Ulas B., Kalomeni B., Keskin V., Kose O., Yakut K., 2011, arXiv, arXiv:1107.0277
\bibitem[\protect\citeauthoryear{van Hamme}{1993}]{ha}van Hamme, W., 1993, AJ, 106, 2096
\bibitem[\protect\citeauthoryear{von Zeipel}{1924}]{ze}von Zeipel, H., 1924, MNRAS, 84, 665
\bibitem[\protect\citeauthoryear{Vukasovi{\'c}}{1997}]{1997IAPPP..67...11V} Vukasovi{\'c} M., 1997, IAPPP, 67, 11
\bibitem[\protect\citeauthoryear{Wilson and Devinney}{1971}]{wi}Wilson, R.E., Devinney, E.J., 1971, ApJ, 166, 605
\bibitem[\protect\citeauthoryear{Wronka et al.}{2010}]{2010AJ....139.1486W} Wronka M.~D., Gold C., Sowell J.~R., Williamon R.~M., 2010, AJ, 139, 1486
\bibitem[\protect\citeauthoryear{Yakut \& Eggleton}{2005}]{2005ApJ...629.1055Y} Yakut K., Eggleton P.~P., 2005, ApJ, 629, 1055
\bibitem[\protect\citeauthoryear{Yakut et al.}{2005}]{2005MNRAS.363.1272Y} Yakut K., Ula{\c s} B., Kalomeni B., G{\"u}lmen {\"O}., 2005, MNRAS, 363, 1272
\bibitem[\protect\citeauthoryear{Zabihinpoor, Dariush, \& Riazi}{2006}]{2006ApSS.302...27Z} Zabihinpoor S.~M., Dariush A., Riazi N., 2006, Ap\&SS, 302, 27
\bibitem[\protect\citeauthoryear{Zhang}{2010}]{2010PASP..122..309Z} Zhang L.-Y., 2010, PASP, 122, 309
\end{thebibliography}
\end{document}